\documentclass[12pt]{emulateapj}

\usepackage{graphicx}
\shortauthors{Pineda et al.}
\shorttitle{}
\begin{document}

\title{THE SLOAN DIGITAL SKY SURVEY DATA RELEASE 7 SPECTROSCOPIC M DWARF CATALOG III: The SPATIAL DEPENDENCE OF  MAGNETIC ACTIVITY IN THE GALAXY}

\author{J. Sebastian Pineda \altaffilmark{1,2}, 
Andrew A. West\altaffilmark{3}, John J. Bochanski \altaffilmark{4,5}, Adam J.
Burgasser \altaffilmark{6}}

\altaffiltext{1}{Corresponding author: jspineda@astro.caltech.edu}
\altaffiltext{2}{California Institute of Technology, Department of Astronomy, 1200 E. California Ave, Pasadena CA, 91125, USA}
\altaffiltext{3}{Department of Astronomy, Boston University, 725 Commonwealth Avenue, Boston, MA 02215, USA}
\altaffiltext{4}{Haverford College, Department of Physics and Astronomy, 370 Lancaster Ave., Haverford, PA, 19041, USA.}
\altaffiltext{5}{Department of Astronomy and Astrophysics, The Pennsylvania State University, 525 Davey Laboratory, University Park, PA 16802, USA}
\altaffiltext{6}{Center of Astrophysics and Space Sciences, Department of Physics, University of California, San Diego, CA 92093, USA}

\begin{abstract} 
We analyze the magnetic activity of 59,318 M dwarfs from the Sloan Digital Sky Survey (SDSS) Data Release 7. This analysis explores the spatial distribution of M dwarf activity as a function of both vertical distance from the Galactic plane $(Z)$ and planar distance from the Galactic center $(R)$. We confirm the established trends of decreasing magnetic activity (as measured by H$\alpha$ emission) with increasing distance from the mid-plane of the disk and find evidence for a trend in Galactocentric radius. We measure a non-zero radial gradient in the activity fraction in our analysis of stars with spectral types dM3 and dM4. The activity fraction increases with $R$ and can be explained by a decreasing mean stellar age with increasing distance from the Galactic center.
\end{abstract}

\section{Introduction}

Low-mass stars such as M dwarfs, have lifetimes that exceed the current age of the Universe. As they constitute the majority of all stars in the Milky Way, they are ideal for examining the evolution and dynamics of the Galaxy (e.g. \citealt{fuchs2009}).  M dwarfs also harbor magnetic dynamos \citep{browning2006}, which are revealed through specific signatures of chromospheric heating (magnetic activity), such as hydrogen Balmer series and Ca {\small II} H and K emission lines \citep{skumanich1972, duncan1991, hawley1996,west2004, west2008, jenkins2011}. These emission features have been linked with stellar rotation and age \citep{soderblom1991, hawley1996, hawley1999, reiners2008}, with younger stars appearing to be more magnetically active than older stars \citep{gizis2002}. The functional relationship between age and magnetic activity is starting to take shape and a coherent picture is emerging. Angular momentum losses, through stellar winds spinning down the star, play a role in the decline of magnetic strength and activity with stellar age \citep{skumanich1972, reiners2008}. Additionally, the rate at which angular momentum is lost changes significantly across the boundary between stars with partially convective envelopes and stars with fully convective envelopes ($\sim$dM4 ; \citealt{shulyak2011, reiners2012}). This is a consequence of the changes in stellar radius across this boundary that lead to slower spin-down rates for the lowest mass stars \citep{reiners2012}. 
 
Recent studies have examined how magnetic activity in M dwarfs varies as a function of spectral type and position in the Galaxy (\citealt{west2006}; \citealt{west2008}; \citealt{west2011}; hereafter Paper I), taking advantage of large photometric and spectroscopic samples from the Sloan Digital Sky Survey (SDSS; \citealt{york2000}; \citealt{ivezic2004}). Using H$\alpha$ emission as a proxy for magnetic activity, \cite{west2008} showed that for each spectral type, magnetic activity decreases as a function of absolute vertical distance away from the Galactic plane, in agreement with kinematic studies suggesting that populations farther from the Galactic mid-plane are likely older. Accordingly both Galactic position and magnetic activity can be used as a proxy for age \citep{wielen1977, west2008, fuchs2009}. \cite{west2008} also estimated magnetic activity lifetimes, showing that they increase with spectral type such that later types (e.g. $\gtrsim$dM4) remain magnetically active longer than early types (e.g. $\lesssim$dM4; \citealt{west2008}).

\cite{west2008} studied trends in activity as a function of one dimension only (the absolute vertical distance from the Galactic mid-plane) smoothing over any potential trends in the radial direction from the underlying density distribution of the disk, the recent star formation history and the relative contributions of disk and halo populations. In the inside-out growth scenario for Milky-way like galaxies, stars of the inner disk could have formed from early episodes of star formation and represent an older stellar population compared to stars farther from the galactic center formed in the more recent past \citep{freeman2002}. Indeed, other disk galaxies have been shown to exhibit radial age gradients \citep{macarthur2004, barker2011}. However, the observed distributions in age may not be due solely to \textit{in situ} star formation, as stellar migration can also play a role. Simulations have shown that the mixing of stellar populations through migration can alter the observed structure of Galactic disks (\citealt{roskar2008}; \citealt{schonrich2009}; \citealt{loebman2011}). Nevertheless, the Geneva Copenhagen Survey (GCS; \citealt{nordstrom2004}; \citealt{casagrande2011}) of F and G dwarfs showed a negative radial age gradient for stars near the plane of the Milky Way disk (mostly younger thin disk stars) of $\sim$1 Gyr kpc$^{-1}$. If the average magnetic activity of a stellar population can be used as a proxy for its age, then the imprint of this observed age gradient might be evident in the stellar magnetic activity distribution in the Galaxy. These results suggest that we should expect to find more active stars at greater Galactocentric distances.

In this article, we report an analysis of the magnetic activity of 59,318 M dwarfs from SDSS DR7 as functions of both the distance away from the Galactic mid-plane and Galactocentric radius. Using the SDSS data, we aim to demonstrate whether or not there is a significant radial trend in the M dwarf activity distribution. In Section~\ref{sec:data} we summarize the publicly available data sample used in the analysis. In Section~\ref{sec:act} we examine the magnetic activity as it varies throughout the Galaxy and quantify its spatial gradient. In Section~\ref{sec:bias} we consider some of the potential systematics that could affect our analysis. Lastly, in Section~\ref{sec:discuss} we summarize and discuss the implications these trends have on the age distribution of stars in the Galaxy.

\begin{figure}[htbp]
\centering
\includegraphics[width=.65\textwidth,clip=true, angle=0]{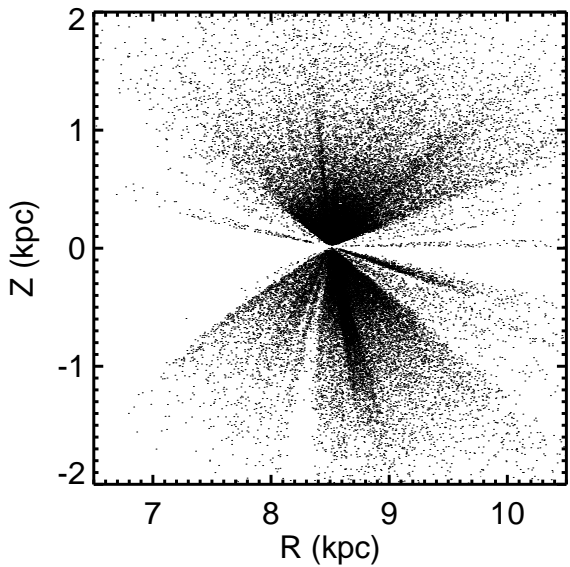}
\caption[DR7 M dwarf Position Map]{Positions for DR7 M dwarf sample plotted in cylindrical Galactocentric coordinates, $R$ and $Z$. Positive $Z$ corresponds to the northern hemisphere. The position of the Sun is taken to be $(R,Z) = (8500,15)$ pc.}
\label{fig:pos}
\end{figure}

\begin{deluxetable*}{ccccccc} \tablewidth{\textwidth} 
\tablecolumns{6}
\tablecaption{M dwarf activity by spectral type}
\tablehead{   
  \colhead{Spectral Type} &
  \colhead{N} &
  \colhead{`Active'} &
  \colhead{`Not-Active'} &
	\colhead{`Weak'} &
	\colhead{Mean Activity \% }&
	\colhead{Activity Lifetimes (Gyr) \tablenotemark{a}}
}
\startdata
dM0 & 9,979 & 222 & 9,497 & 260 & $2.3\pm0.2$& $0.8 \pm 0.6$\\ 
dM1 & 8,199 & 231 & 7,594 & 374 & $3.0 \pm 0.2$&$0.4 \pm 0.4$\\
dM2 & 9,367 & 358 & 8,444 & 565 & $4.1\pm 0.2$&$1.2 \pm 0.4$\\
dM3 & 10,068 & 605 & 8,561 & 902 &$ 6.6 \pm 0.3$& $2.0 \pm 0.5$\\
dM4 & 7,985 & 1,006 & 5,909 & 1,070 & $14.5 \pm 0.4$ &$4.5^{+0.5}_{-1.0}$ \\ 
dM5 & 3,483 & 1,083 & 1,715 & 685 & $38.7 \pm 0.9$& $ 7.0 \pm 0.5$\\
dM6 & 4,994 & 2,152 & 1,618 & 1,224 & $57.8\pm 0.8$&$7.0 \pm0.5$\\
dM7 & 4,447 & 2,047 & 1,114 & 1,286 & $64.8 \pm 0.8$&$8.0^{+0.5}_{-1.0}$\\
dM8 & 600   &  330  &   80  &   190    &  $80.5 \pm 2.0 $& -- \\
dM9 &  196   & 114    &  22 &   60        &  $83.8^{+2.6}_{-3.6} $& --
\enddata
\label{tab_act}
\tablenotetext{a}{Liftetimes from \cite{west2008}; they did not determine lifetimes for dM8 or dM9}
\end{deluxetable*}

\section{Data}\label{sec:data}

We used a sample of M dwarfs from the SDSS DR7 that contains 70,841 stars with observed spectra and photometry (Paper I). The selection criteria were based on previous M dwarf samples \citep{west2008, kowalski2009}. For spectral classification, we used the HAMMER \citep{covey2007} spectral typing software to visually inspect each spectrum (amongst 17 individuals). Paper I demonstrated that the automatic spectral types from the HAMMER are accurate to within $\pm1$ sub-type for early M dwarfs ($<$M5) but systematically off by 1 subtype for late-type M dwarfs ($\ge$M5). Thus, the assigned spectral types were determined manually by the various individuals and a control sample was established to test the reliability of the visual inspections. This test showed that the dispersion of the classifications for each star was $<0.4$ subtypes with no individual producing spectral types that were systematically off by more than $0.2$ subtypes (see Paper I for further details). The visual inspections also allowed us to cull out any contaminants such as giant stars or subdwarfs.

We used SDSS photometric flags  to apply additional cuts to ensure a high quality data sample (see Paper I). We then reduced the full sample to $N = 59,318$ by only considering those stars with a signal-to-noise ratio (S/N) $> 3$ in the continuum near H$\alpha$. We determined the distances to all of the stars using the $M_r$ vs. $r - z$ photometric parallax relation of \cite{bochanski2010}, where we have accounted for the effects of dust and reddening \citep{jones2011}.

Using the photometric distances and SDSS astrometry, we estimated a spatial position ($R$ and $Z$ in Galactocentric cylindrical coordinates) in the Galaxy for each star assuming that the Sun is $15$ pc above the mid-plane of the Galactic disk (\citealt{binney1997}) and $8.5$ kpc from the center of the Galaxy (\citealt{kerr1986}). In Figure~\ref{fig:pos}, we plot the positions of the stars in our sample.

For each star we calculated the equivalent width (EW) of H$\alpha$ as a measure of magnetic activity (\citealt{hawley1996}; \citealt{walkowicz2004}). We classified stars as `active' according to the following criteria:

\begin{itemize}
\item EW in H$\alpha$ emission $>0.75$ \r{A}
\item EW $>$ 3 times the uncertainty in EW
\item H$\alpha$ emission line height $>$  3 times the continuum level
\item S/N $>$ 3 near H$\alpha$
\end{itemize}

\noindent Those stars that did not satisfy the first three of these criteria but with S/N $>$ 3 near H$\alpha$ were labeled as inactive (see Paper I for more details). \cite{west2008} also performed a Monte Carlo simulation in which they demonstrated that these criterion allowed them to distinguish between active and inactive stars without introducing a distance-related bias to the activity designation. Our final sample includes 8,149 active stars and 44,554 inactive stars. The remaining stars were considered to have `weak' activity or low S/N and were excluded from the analysis as inconclusive. The sample also included 796 stars of spectral types dM8 or dM9 that were not included in the spatial analysis due to the small sample size. However, we did measure the average activity level of these stars as shown in Table~\ref{tab_act}.

The sample used in this study does not represent a complete selection of M dwarfs in the local solar neighborhood. However, the sample spans a wide and representative range of stellar properties (see Paper I). Accordingly, there may be a variety of effects related to changes in the inferred luminosity of a given star that could affect the observed distribution of stellar properties. These include deviations from the nominal photometric parallax relation due to metallicity effects and/or a difference in the intrinsic luminosity of active stars versus inactive stars. We investigate potential systematics due to these effects in Section~\ref{sec:bias}.

 \begin{figure*}[htbp]
\centering
\includegraphics[clip=true,angle=0,width=\textwidth]{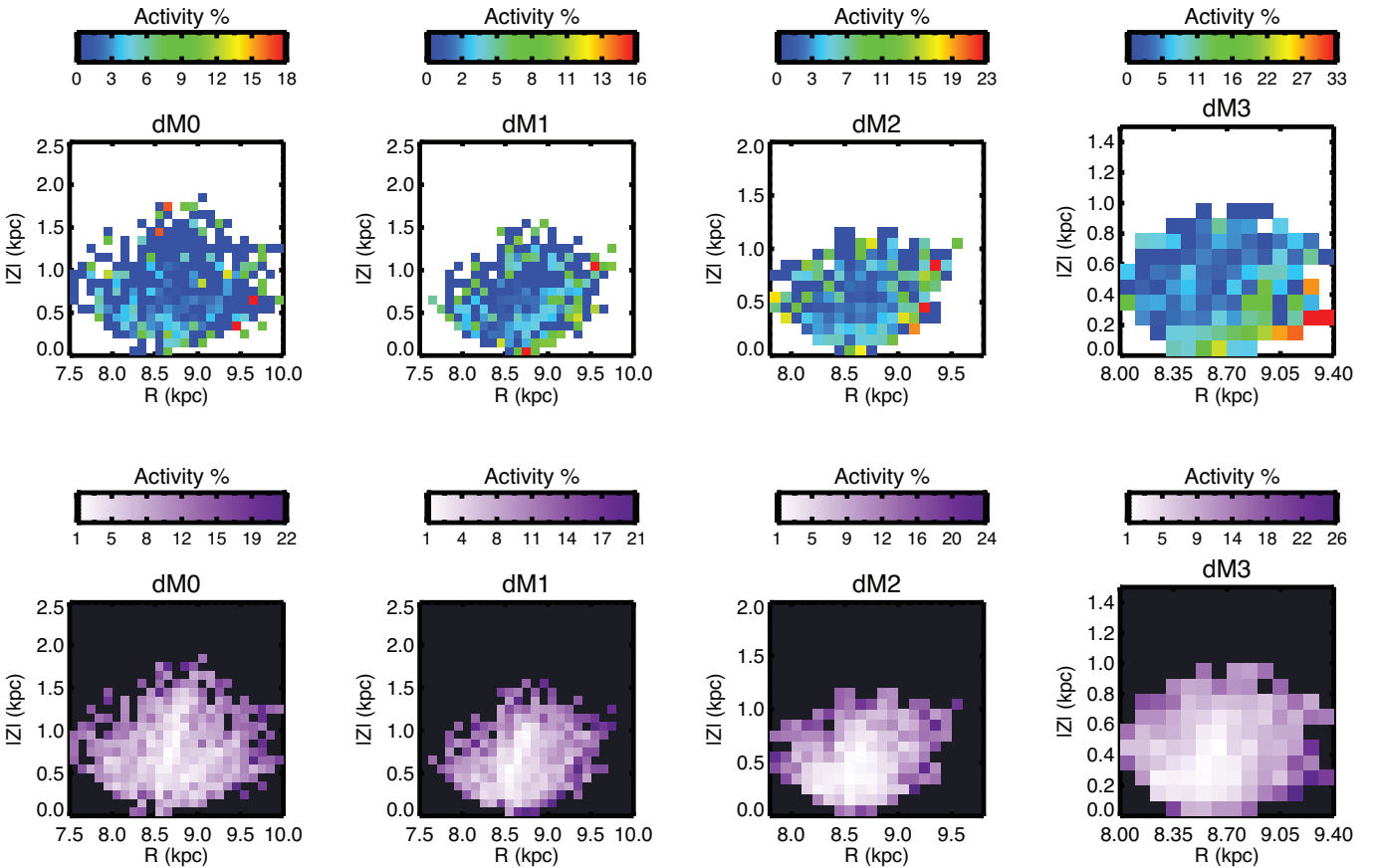}
\caption[]{ Top - Activity fraction maps for spectral types dM0 - dM3, as functions of Galactocentric radius and absolute distance from the Galactic mid-plane. The color corresponds to the activity fraction, defined as the number of active stars divided by the sum of the number of active and not-active stars. Redder colors correspond to bins that are more active. Stars are separated into 100 pc by 100 pc, bins. Bottom - Uncertainty pertaining to the activity fraction maps directly above, plotting the total length of the 68\% confidence interval as derived from the binomial distribution.  Dark (violet) shades correspond to bins that have greater uncertainty. The map for dM3 (far right) suggests a possible radial trend in activity. }
\label{fig:frac8}
\end{figure*}

\begin{figure*}[htbp]
\centering
\includegraphics[clip=true,angle=0,width=\textwidth]{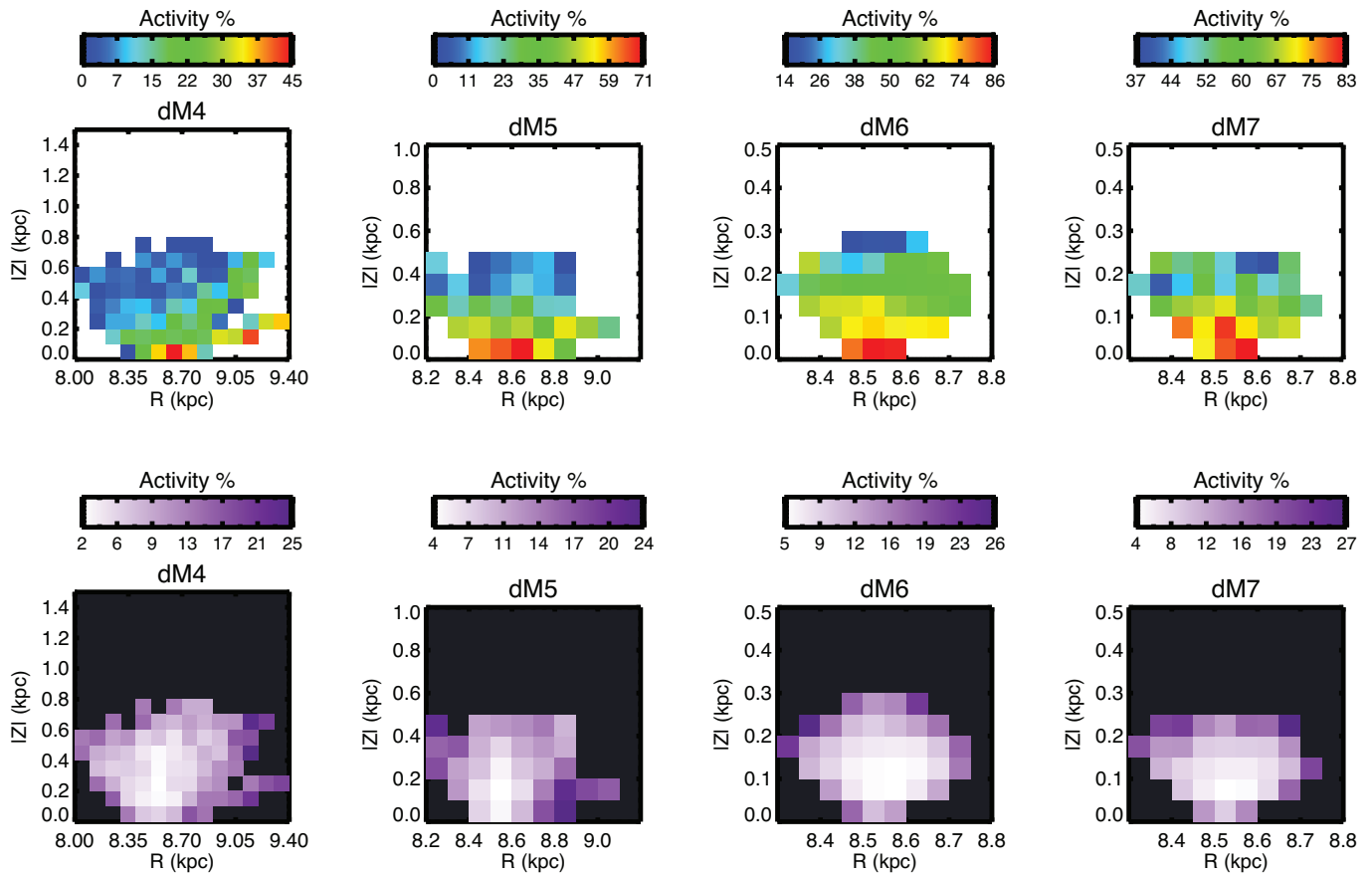}
\caption[]{ Same as Figure~\ref{fig:frac8} but for spectral types dM4--dM7. Spectral types dM6 and dM7 are separated into 50 pc by 50 pc bins. There are strong vertical gradients apparent in these late-type M dwarfs. The map for dM4 (far left) suggests a possible radial trend in activity.}
\label{fig:frac47}
\end{figure*}

\section{Activity Fractions}\label{sec:act}

We first separated the stars according to their spectral type as shown in Table~\ref{tab_act}. The mean activity \%, $N_{\rm{active}} / (N_{\rm{active}} + N_{\rm{notactive}})$, gives an overall measure of activity for each spectral type in the sample. In agreement with previous findings (\citealt{west2008}), the level of activity rises for later spectral types with a jump in activity at the boundary between stars with a partially convective envelope ($\lesssim$ dM4) and stars with a fully convective envelope ($>$ dM4). Using dynamical simulations, \cite{west2008} determined activity lifetimes for each of these spectral types (see Table~\ref{tab_act}). The observed stellar activity level therefore depends on the type of star and the mean stellar age, which varies as a function of position in the Galaxy (\citealt{loebman2011}).

\subsection{Maps}\label{sec:map}

Examining each spectral type individually and folding the data across the mid-plane (assuming symmetry), we separated the stars into ($R$, $Z$) bins and calculated the fraction of stars in each bin that were active. In Figure~\ref{fig:frac8} and Figure~\ref{fig:frac47} we show the distributions for spectral types dM0 - dM7. The color in the maps of the top row correspond to the activity fraction, where redder indicates a larger active fraction. For all of the maps, except those for dM6 and dM7 dwarfs, we used a binning of $100$ pc by $100$ pc to make sure that each bin has a significant number of stars ($\ge$10) and to reduce the uncertainty associated with the fraction determination. For spectral types dM6 and dM7 we used a binning of $50$ pc by $50$ pc because of the large number of late-type M dwarfs near the solar neighborhood (9,194 stars within 300 pc). In the bottom row of Figure~\ref{fig:frac8} and Figure~\ref{fig:frac47} we plot the corresponding uncertainty maps showing the length of the 68\% confidence interval associated with each fraction determination calculated using binomial statistics. Because the binomial distribution is asymmetric, the uncertainties are not symmetric across the median values. 

The maps (Figure~\ref{fig:frac8} and Figure~\ref{fig:frac47}) reproduce the previous results of \cite{west2008} with regard to the decreasing trend in activity away from the Galactic plane, interpreted as a consequence of a vertical age gradient. The trend is evident in all but the plots for the early-type M dwarfs (dM0--dM1) due to the small fraction of active stars. For the mid-type M dwarfs (dM3--dM4) it appears that there may be a possible increase in activity away from the Galactic center. The late-type stars (dM5--dM7) do not appear to show a radial trend either. The early types are not sufficiently active for there to be a noticeable trend, whereas for the late-type stars a gradient in activity may not be sufficiently large across the sample to be noticeable at the current ages of these stars. Only for the mid-type M dwarfs is the gradient steep enough in the data. The maps are suggestive and we quantify the significance of any possible trends in Section~\ref{sec:grad}.

\subsection{Gradients}\label{sec:grad}

Using the activity fraction as a function of position we quantified the trends shown in the maps. We modeled the activity fraction as,

\begin{equation}
 \mathcal{F}  = a |Z| + b(R - R_{\odot}) + c \; ,
 \label{eq:mod}
\end{equation}

\noindent where the positions, $Z$ and $R$ are in kpc and we subtracted by $R_{\odot} = 8.5$ kpc (\citealt{kerr1986}) radially for convenience as our data are situated about the position of the Sun. The parameters $a$ and $b$ describe the gradients in the $Z$ and $R$ directions respectively whereas $c$ is closely related to the mean activity fraction. There is no clear functional relationship between the activity fraction and Galactic position however, this model choice allowed for a simple quantitative test to assess the significance of trends that may or may not be visually apparent in the maps of Figure~\ref{fig:frac8} and Figure~\ref{fig:frac47}.

The joint probability distribution for the parameter fits for each map was given by:

\begin{equation}
p(\{abc\}; \{data\}) \propto p(\{data\} ; \{abc\} ) p( \{abc\}  ) \; ,
\label{eq:prob}
\end{equation}

\noindent where the first term on the right represents the likelihood of the data and the second term on the right is the prior distribution on the parameters, which we took to be uninformative. Each data point corresponds to a spatial bin with $k_i$ active stars in $N_i$ total stars. The likelihood is the product of the probabilities for each of these data points given an underlying activity fraction distribution modeled by Equation \ref{eq:mod}. Thus, using binomial statistics the probability of each datum is given by

\begin{eqnarray}
p( k_{i}; \{N,R,|Z|\}_i ,\{abc\})=&  \nonumber \\
  \frac{N_i!}{k_i!(N_i-k_i)!}& \mathcal{F}^{k_{i}} (1 - \mathcal{F})^{N_{i}-k_{i}} \; ,
\end{eqnarray}

\noindent where $\mathcal{F}(a,b,c)$ is given by Equation \ref{eq:mod}. The index $i$ runs over all the bins with at least one star. The high uncertainty in the less populated bins is naturally taken into account by the probability theory. We explored the joint probability distribution of Equation \ref{eq:prob} using Markov Chain Monte Carlo (MCMC) methods while restricting the values of $\mathcal{F}$ as predicted by the model to the physical range of 0 to 1. The best fit parameters are then determined from the medians of the marginalized distributions. 

In Table~\ref{tab:fit} we show the results of this analysis where the error bars indicate the 99.73\% confidence interval (3$\sigma$). Our estimates for the vertical gradients are shown in column $a$. For spectral types dM0--dM2, the vertical gradients are small and not significant. However, for spectral types dM3--dM7 they are stronger and significant at greater than the $3\sigma$ level. Our estimates for the radial gradients are shown in column $b$. For spectral type dM0, this gradient is consistent with zero. For spectral types dM1, dM2 and dM4, the radial gradients are inconsistent with zero at the $2\sigma$ confidence level and for spectral type dM3, inconsistent with zero beyond the $3\sigma$ confidence level. After the convective boundary ($\gtrsim$ dM4) the local activity level (column $c$) jumps up and the radial gradients appear to reverse direction although they are consistent with zero. Additionally, this trend may be spurious as stars of spectral types dM5--dM7 do not probe far in the radial direction. These gradients are only applicable to the regions shown on the maps in Figure~\ref{fig:frac8} and Figure~\ref{fig:frac47} with values of $\mathcal{F}$ between 0 and 1. 

 \begin{figure}[htbp]
\centering
\includegraphics[scale=.5,clip=true,angle=0]{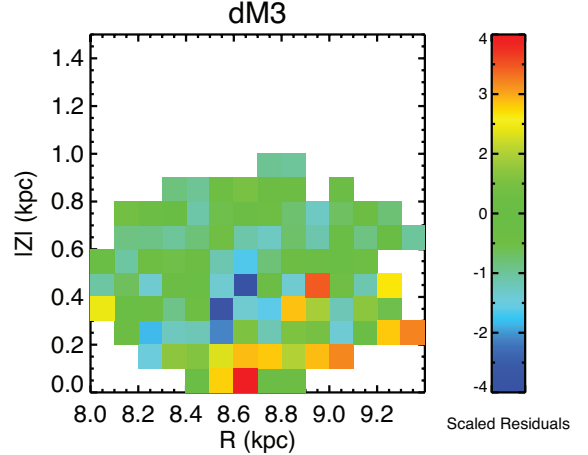}
\caption[]{ A comparison of the model prediction and the data for stars of spectral type dM3. The scaled residuals (Equation \ref{eq:res_m3}) show the difference between the model prediction and the data scaled by the square-root of the variance of the binomial statistic. In the limit of there being a large number of stars in a given bin this scaling is equivalent to scaling by the standard deviation of a normal distribution, about half of its full width at half maximum. The positive residuals near the mid-plane indicate that the linear model in the vertical direction is insufficient to describe the drop in activity as a function of height above the mid-plane.}
\label{fig:res_m3}
\end{figure}

In Figure~\ref{fig:res_m3} we compare the model with the data for dM3 stars, showing the scaled residual for each bin, given by

\begin{equation}
Res_i = \frac{k_{i} - N_i \mathcal{F}}{\sqrt{N_i \mathcal{F} (1 - \mathcal{F})}} \; ,
\label{eq:res_m3}
\end{equation}

\noindent where $k_i$ are the number of active stars in the bin, $N_i$ is the total number of stars in that bin and $\mathcal{F}$ is the activity fraction of Equation \ref{eq:mod}. The denominator represents the typical half-width of a binomial distribution with probability $\mathcal{F}$ and $N$ attempts. Thus, positive values of the scaled residuals mean that the model is under predicting the number of active stars whereas negative values mean the model is over-predicting the number of active stars.  From this plot it is evident that although the model broadly reproduces the data it under-predicts the number of active stars near the mid-plane. This suggests that the activity-fraction may drop off rapidly at low $|Z|$ and then level off with increasing absolute vertical distance from the mid-plane. There does not appear to be a striking deviation in the radial direction from our linear model. Although the model averages over the vertical behavior it appears sufficient to track trends in the radial direction. Future studies may assess the functional form of the activity fraction distribution, however that is beyond the scope of the present study.

\begin{deluxetable}{cccc}  
\tablecolumns{4}
\tablecaption{Activity Fraction Gradients\tablenotemark{i}}
\tablehead{   
  \colhead{Spectral Type} &
  \colhead{$a$ (kpc$^{-1}$)} &
  \colhead{$b$ (kpc$^{-1}$)} &
  \colhead{$c$} 
}
\startdata
dM0 & $-0.007\pm^{0.009}_{0.008}$& $0.002\pm 0.008$  &$ 0.025 \pm 0.009$\\[4pt] 
dM1 & $0.001 \pm^{0.014}_{0.013}$& $0.009 \pm^{0.010}_{0.011}$ & $0.026 \pm^{0.012}_{0.010}$\\[4pt]
dM2 & $-0.015\pm_{0.011}^{0.016}$&$0.012 \pm^{0.014}_{0.013}$&$0.045 \pm^{0.011}_{0.010}$\\[4pt]
dM3 & $-0.028\pm_{0.009}^{0.015}$ & $0.035 \pm^{0.017}_{0.020}$&$0.073 \pm0.010$\\[4pt]
dM4 & $-0.080\pm_{0.012}^{0.014}$& $0.036\pm_{0.038}^{0.022}$&$0.167\pm^{0.017}_{0.015}$\\ [4pt]
dM5 & $-0.240\pm_{0.018}^{0.033}$&$-0.066\pm_{0.067}^{0.096}$&$0.450 \pm_{0.032}^{0.030}$ \\[4pt]
dM6 &$-0.590\pm_{0.050}^{0.097}$ &$-0.079\pm^{0.123}_{0.202}$ &$0.664 \pm_{0.031}^{0.029}$\\[4pt]
dM7 & $-1.108\pm^{0.241}_{0.191}$& $-0.244\pm^{0.285}_{0.282}$&$0.792 \pm^{0.040}_{0.044}$
\enddata
\label{tab:fit}
\tablenotetext{i}{Parameters from model: $\mathcal{F}  = a |Z| + b(R - R_{\odot}) + c$  ; \\ error bars correspond to 99.73\% confidence interval ($3\sigma$)}
\end{deluxetable}

\subsection{Asymmetries}

Although we did not expect significant asymmetries, we checked for them by comparing the gradients in activity fraction in the northern celestial hemisphere with stars in the southern celestial hemisphere (above and below the Galactic plane, respectively). Using the same analysis of Section~\ref{sec:grad}, we estimated the gradients in the two regions; the resulting values are shown in Table~\ref{tab:gradcomp} with 3$\sigma$ uncertainties. Each column lists the difference in the linear coefficients in the sense of north minus south. A value of zero in these columns indicates complete symmetry. Negative values in these columns indicate that the gradient estimates ($a$, $b$) below the plane were larger in absolute value than above the plane or that the typical activity level ($c$) was larger below the plane than above it.

\begin{figure}[htbp]
\centering
\includegraphics[scale=.5,clip=true,angle=0]{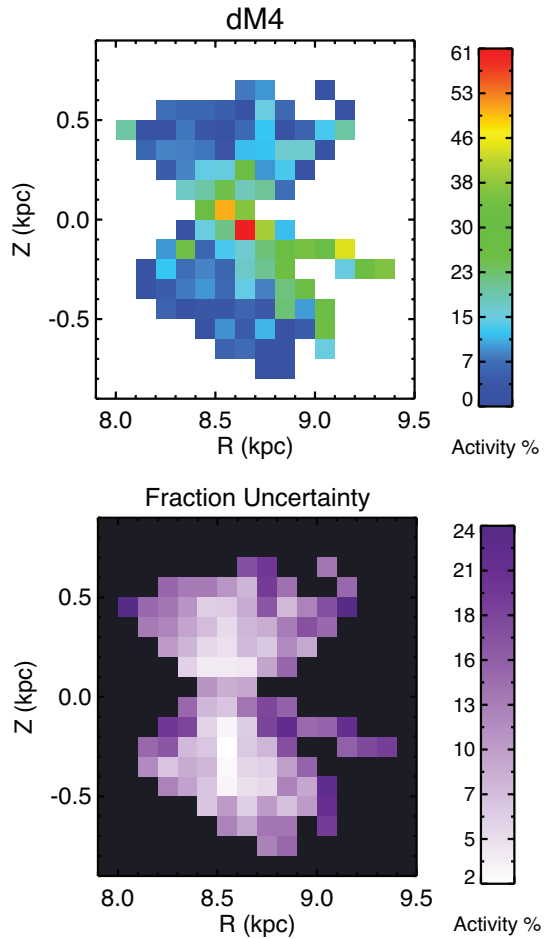}
\caption[]{ Top - dM4 activity fractions as functions of Galactocentric radius and distance from the Galactic plane. Redder colors correspond to bins that are more active. Bottom - Map of total uncertainties corresponding to the activity fraction map of the top panel. Uncertainties determined from the binomial distribution. Lighter shades correspond to lower uncertainty. Bins are 100 pc by 100 pc. Shows the activity fractions for both above and below the plane and suggests that there may be an asymmetry, likely due to differences in the sampling above and below the mid-plane}
\label{fig:m4_act_frac2}
\end{figure}

For the stars of spectral type dM0--dM2, the gradients are rather small and mostly in accord. The values of $\Delta c$ for these types are consistent with zero. For stars of spectral type dM3--dM5 the differences in the vertical gradients are close to zero but the radial gradients vary. From the negative values of $\Delta b$ in the analysis of dM3, dM4, it is evident that the radial trends are dominated by stars below the Galactic plane (see Figure~\ref{fig:m4_act_frac2}). For dM5 and dM7 stars both regions were consistent with each other. For dM6 stars the large gradient difference is due to an apparent reversal below the plane. However for dM6 stars the radial gradient was not strong and the two hemispheres are generally consistent.

There does however appear to be an asymmetry between the north and south in the dM4 stars, perhaps related to sampling effects. The north and south are sampled differently, including a southern sightline that is not matched well in the northern hemisphere. Any differences shown in Table~\ref{tab:gradcomp} may be due to this disparity. As the northern sample is more densely sampled overall ($29\%$ more stars) than the southern sample and since over all of the spectral types the hemispheres do not differ much, we complete the rest of our analysis merging the two hemispheres to improve our statistics.

\begin{deluxetable}{cccc}  
\tablecolumns{4}
\tablecaption{Asymmetry in Activity Gradients \tablenotemark{i}}
\tablehead{   
  \colhead{Spectral Type} &
  \colhead{ $\Delta a$ \tablenotemark{ii} (kpc$^{-1}$)} &
  \colhead{ $\Delta b$ \tablenotemark{iii} (kpc$^{-1}$)} &
  \colhead{$\Delta c$ \tablenotemark{iv}  (kpc$^{-1}$) } 
}
\startdata
dM0 & $-0.009\pm_{0.017}^{0.016} $ & $0.000\pm_{0.016}^{0.017}$  &$ 0.005 \pm0.018$\\ [4pt]
dM1 & $ -0.005\pm_{0.025}^{0.026}$& $-0.016 \pm_{0.021}^{0.022}$ & $-0.001 \pm_{0.021}^{0.023}$\\ [4pt]
dM2 & $0.019\pm_{0.026}^{0.031}$&$-0.019\pm_{0.023}^{0.025} $&$0.016\pm_{0.020}^{0.021} $\\ [4pt]
dM3 & $0.016 \pm^{0.033}_{0.029}$& $-0.030\pm_{0.034}^{0.039}$&$ 0.019 \pm_{0.023}^{0.021}$\\ [4pt]
dM4 & $0.025 \pm^{0.052}_{0.048}$& $-0.110\pm_{0.077}^{0.070}$&$0.060 \pm 0.036$\\  [4pt]
dM5 & $0.022\pm_{0.060}^{0.086}$&$0.003\pm_{0.128}^{0.227}$&$0.180\pm_{0.063}^{0.065}$ \\ [4pt]
dM6 &$-0.749\pm^{0.776}_{0.436}$ &$0.416\pm_{0.572}^{0.435}$ &$-0.058\pm_{0.109}^{0.131}$\\ [4pt]
dM7 & $-0.495 \pm^{1.642}_{1.572}$& $0.026\pm_{1.229}^{1.230}$&$-0.085 \pm^{0.169}_{0.148}$
\enddata
\label{tab:gradcomp} 
 \tablenotetext{i}{Uncertainties represent 99.73\% confidence interval (3$\sigma$)}
 \tablenotetext{ii}{$\Delta_{a} = -a_{\rm{North}} - a_{\rm{South}}$}
  \tablenotetext{iii}{$\Delta_{b} = b_{\rm{North}} - b_{\rm{South}}$}
  \tablenotetext{iv}{$\Delta c = c_{\rm{North}} - c_{\rm{South}}$}
\end{deluxetable}

\section{Potential Systematics}\label{sec:bias}

\subsection{Spectral Misclassification}

Our activity maps vary as a function of spectral type. Although the spectral types of our stars are accurate to about 0.5 subtypes (see Section~\ref{sec:data}), it is possible that spectral misclassification could influence our measurement of the activity fraction distribution in the Galaxy. To investigate the possibility of spectral misclassification, we examined what the effect would be on the activity distribution by contamination from adjacent spectral types. We considered stars within a standard deviation of the median $r - z$ color of the given spectral type to have likely been assigned their types correctly. We then took the set of stars that fell between the median $r - z$ color of the adjacent spectral types and one standard deviation from the median color as potential candidates for inclusion in the contaminated sample. From this set of stars we randomly selected stars to reclassify as the central spectral type until we matched the original sample size.

In practice, when simulating this contamination for 10,068 dM3 stars, about 16\% are switched with stars of similar colors, with half of those stars classified as dM2 and the other half classified as dM4. For dM4 stars, the simulated contamination contained 9\% dM3 stars and 5\% dM5 stars; the disparity being a consequence of there being many more dM4 and dM3 stars than dM5 stars in the sample. By keeping the total number constant we maintained the statistics of the original classification, allowing for a clearer comparison between the original activity maps and the new ``contaminated" maps. This procedure mimics a sample selected by their $r-z$ color, with bins centered around the median color of the desired spectral type.

\begin{figure*}[htbp]
\centering
\includegraphics[width=\textwidth]{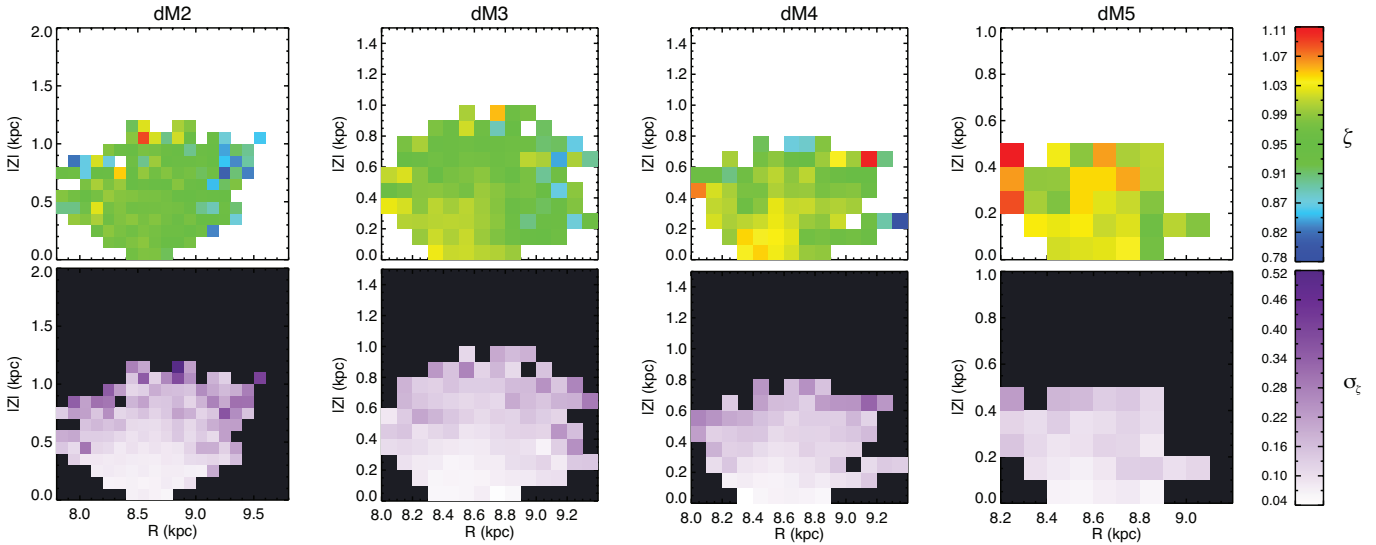}
\caption[]{Maps showing the distribution of the metallicity proxy, $\zeta$ for spectral types dM2-dM5. Along the top row we plot the median $\zeta$ value for each bin (100 pc by 100 pc) and in the bottom row the corresponding scatter. Thin disk stars have a typical zeta value of $\zeta \sim1$. There appears to be a slight gradient in the metallicity which could effect the measured activity distribution; we examine this effect in Section~\ref{sec:meta}.}
\label{fig:zeta}
\end{figure*}

Using the ``contaminated" sample, we applied our model to measure the gradients and compared them to the original gradient measures, focusing on the types dM3 and dM4 where we see hints of a radial gradient in the activity distribution. In over 100 realizations of the contamination, we measured mean values for the reselected dM3 stars of $a = -0.029\pm0.007$ kpc$^{-1}$, $b = 0.031\pm0.013$ kpc$^{-1}$, and $c = 0.073 \pm0.005$. For the reselected dM4 stars we measured values of $a = -0.080\pm0.017$ kpc$^{-1}$, $b = 0.035\pm0.026$ kpc$^{-1}$, and $c = 0.160 \pm0.008$, where all of the uncertainties here reflect three times the standard deviation of all the values across the many realizations. Although the exact values depend on the randomization, the spread gives the expected range of central values for the gradient measurements due to spectral misclassification with the actual uncertainty in the fit parameter comparable to that given by the uncontaminated original sample (see Table~\ref{tab:fit}). Comparing these values to those found in Table~\ref{tab:fit}, it is evident that the reclassification procedure can weaken the radial gradient somewhat but never eliminate it.  

\subsection{Systematic Spectral Type Shift}

In addition to this statistical misclassification, we considered a systematic misclassification of the sample stars as a function of the distance to each star. This kind of bias could arise if distant stars that were predominately metal-poor compared to nearby stars were misclassified as being an earlier spectral type due to the weakening of metal sensitive absorption features. It is instructive to illustrate the effect this kind of systematic could have on our measured activity distributions. If distant stars are systematically of an earlier type than we can correct this bias by adjusting the spectral type of distant stars to be later than originally assumed. To apply this effect we considered a linear shift in the spectral type with distance.

\begin{equation}
SpT_{new} = SpT_{old} + d*( 1 \, \mathrm{kpc}^{-1}) \; ,
\end{equation}

\noindent where $d$ is the distance to the star. We then rounded to the nearest spectral type to get the adjusted classifications. We then measured the corresponding gradients. For dM3 stars we estimated values of $a = -0.026\pm_{0.009}^{0.012}$ kpc$^{-1}$, $b = 0.011\pm0.012$ kpc$^{-1}$, and $c = 0.058 \pm0.010$. For dM4 stars we estimated values of $a = -0.075\pm^{0.008}_{0.009}$ kpc$^{-1}$, $b = 0.001\pm_{0.006}^{0.009}$ kpc$^{-1}$, and $c = 0.150 \pm_{0.013}^{0.014}$, where the uncertainties reflect $3\sigma$ confidence intervals. In the dM3 stars the gradient is significant at only the $2\sigma$ level and the gradient in dM4 stars disappears. In the case of the dM3 stars the gradient is driven by nearby active stars and the signal is diminished by the incorporation of many inactive stars originally classified as dM2. The weaker dM4 gradient however vanished with this effect.

We showed that this kind of systematic bias in the spectral typing can weaken the observed gradients. However, as applied here the degree of shift was determined arbitrarily without a good observational basis. An additional problem with this distance based shift is that if it mimics a metallicity bias in the spectral typing, stars at lower galactic latitudes are shifted as much as distant stars toward the north galactic pole even though their metallicities may not be as metal-poor. These effects suggest that the systematic shift of spectral types with distance as applied here may be an over correction. Additionally, the spectral types that we determined for our sample are based on the whole optical spectrum of each star  (see Section~\ref{sec:data}) and are less sensitive to changes in metal sensitive absorption features, so this kind of spectral type bias should not be important. We consider additional metallicity effects in the next section.

\subsection{Metallicity-Luminsoity Effects}\label{sec:meta}

Stars of a given intrinsic luminosity have different observed colors depending on the metal content of each star. Since we used the $M_r$ vs. $r - z$ photometric parallax relation of \cite{bochanski2010} to determine the distances to stars, deviations from this relation due to differences in the metallicity would bias our activity distributions, especially if there is a significant underlying metallicity gradient. To examine these effects we looked at the distribution of the metallicity proxy, $\zeta$, as defined by \cite{lepine2007} and updated by \cite{dhital2012}. $\zeta$ is designed to distinguish between main sequence low-mass dwarfs ($\zeta > 0.825$) and metal poor subdwarfs ($\zeta < 0.825$) . Local disk stars have a value of $\zeta\sim1$ and although there is a large amount of scatter, higher values of $\zeta$ generally correspond to higher values of [Fe/H] \citep{woolf2009}.

In Figure~\ref{fig:zeta} we plot the distribution of median $\zeta$ values for the sample and the accompanying spread in each of the four spectral types from dM2--dM5. The metallicity distributions for dM3 and dM4 stars, in which we see a potential activity gradient in the radial direction, do not differ substantially from the distributions of the adjacent-type dM2 and dM5 stars in which we do not see evidence for this activity gradient. There does however appear to be a slight $\zeta$ gradient going from left to right in each of the maps. Although the spread of $\zeta$ for each bin is pretty broad, this could be a potential systematic effect underlying the activity distributions of Section~\ref{sec:act}. 

In \cite{bochanski2011} (hereafter Paper II), the authors showed that at a fixed $r-z$ color or spectral type there was a difference of about 0.6 magnitudes between stars with $\zeta = 1.1$ and $\zeta=0.88$. At a fixed color, the metal poor stars had a fainter absolute magnitude than the metal rich stars. The photometric parallax relation of \cite{bochanski2010} was developed using local disk stars and is therefore predominately applicable to stars around $\zeta\sim1$. Although, our sample is dominated by these kinds of stars, with a median $\zeta$ of 0.97, there are some dwarf stars with $\zeta \ge1.1$ or $\zeta \le 0.88$. Together, these constitute about 25\% of dM3 stars and about 26\% of dM4 stars. 

To investigate the metallicity effects on the activity distributions of dM3 and dM4 stars, we incorporated a metallicity correction when determining the distances to the stars. We corrected the assumed absolute magnitude, $M_r$, from the photometric parallax relation \citep{bochanski2010}, for stars with $\zeta \ge1.1$ or $\zeta \le 0.88$. We increased $M_r$ by 0.3 for the lower metallicity stars and decreased it by 0.3 for higher metallicity stars (made the metal poor stars intrinsically fainter and the metal rich stars intrinsically brighter). We then measured the corresponding gradients. For dM3 stars we estimated values of $a = -0.030\pm_{0.010}^{0.017}$ kpc$^{-1}$, $b = 0.026\pm0.019$ kpc$^{-1}$, and $c = 0.075 \pm0.011$.  For dM4 stars we estimated values of $a = -0.097\pm^{0.018}_{0.011}$ kpc$^{-1}$, $b = 0.011\pm_{0.022}^{0.021}$ kpc$^{-1}$, and $c = 0.175 \pm_{0.016}^{0.016}$, where the uncertainties reflect $3\sigma$ confidence intervals. 

Compared to the gradients of Table~\ref{tab:fit} there is a weakening of the radial gradient. Given the distribution of $\zeta$ as shown in Figure~\ref{fig:zeta}, the distance correction due to metallicity effects preferentially pushes more metal rich stars on the left of those maps to farther distances than similar stars on the right. This consequently pushes active stars farther out on the left more than on the right, negating some of the underlying activity distribution radial gradient. Nevertheless, the dM3 stars persisted in showing a significant radial gradient in their activity distribution. The correction applied to the dM4 stars, however did reduce the significance of that radial gradient to only a 1$\sigma$ deviation. This is similar to the effects of the spectral type shifting but not as drastic. There is however a caveat when considering this kind of correction for the underlying spatial distribution of $\zeta$: M dwarf metallicities are difficult to determine and although, $\zeta$ gives a rough estimate of metallicity (i.e. low vs. high), it does not track metallicity perfectly.  In fact, the spread in $\zeta$ is quite large for disk stars such that our metallicity corrections may introduce significant scatter into the underlying activity distribution.

\begin{figure}[htbp]
\centering
\includegraphics[width=0.46\textwidth]{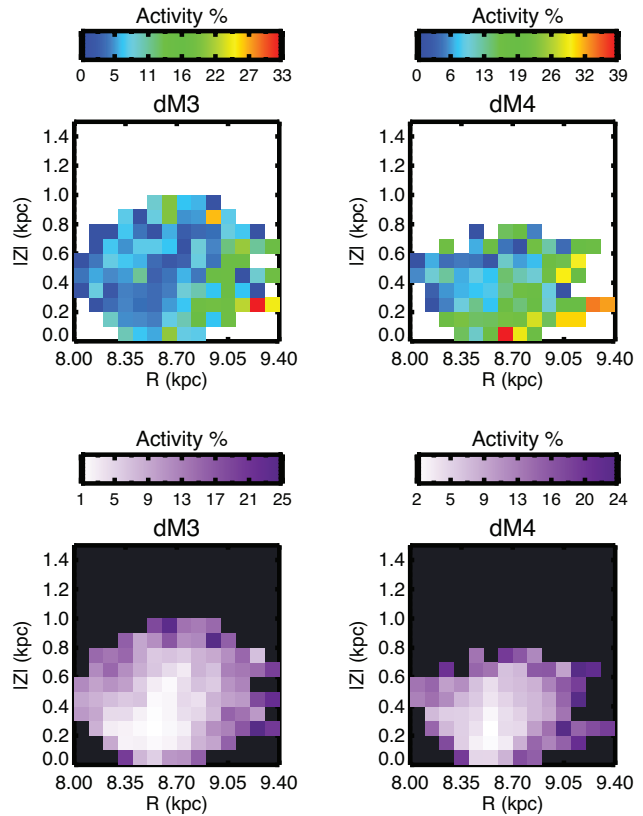}
\caption[]{Activity fraction maps for dM3 and dM4 stars as in Figure~\ref{fig:frac8} and Figure~\ref{fig:frac47} except with adjusted distances for the active stars (see Section~\ref{sec:dist}). Note the enhancement of the radial trend as active stars are pushed to farther distances.}
\label{fig:actdist}
\end{figure}

\begin{figure}[htbp]
\centering
\includegraphics[width=0.46\textwidth]{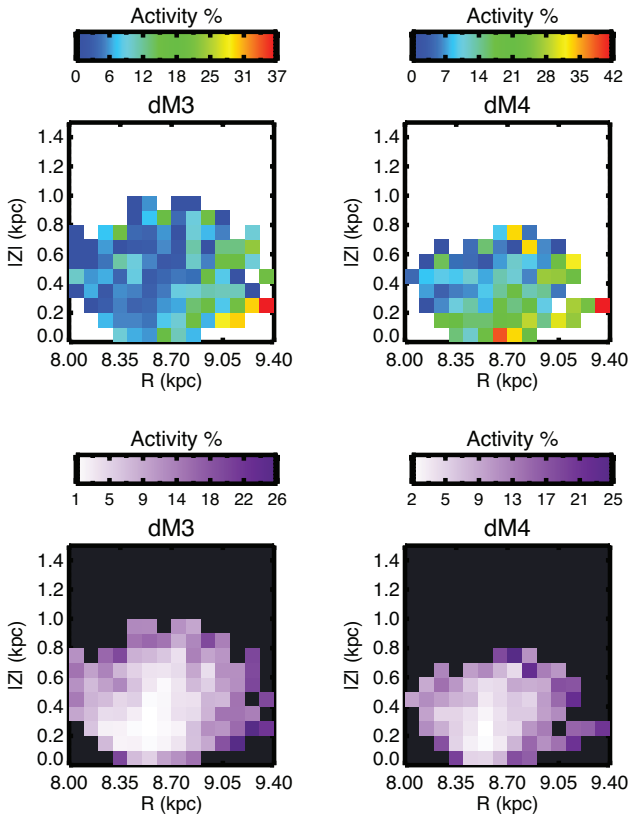}
\caption[]{Activity fraction maps for dM3 and dM4 stars as in Figure~\ref{fig:frac8} and Figure~\ref{fig:frac47} except with distances corrected for activity \emph{and} metallicity effects (see Section~\ref{sec:dist} and Section~\ref{sec:meta}). The effect of the activity distance correction dominates over the metallicity correction.}
\label{fig:actzetadist}
\end{figure}

\subsection{Activity Affecting Distance}\label{sec:dist}

Magnetic activity affects many of the observable properties of low-mass stars \citep{morales2008, morales2010, boyajian2012}. Paper II showed that active stars could be brighter in $M_r$ by as much as half a magnitude compared to inactive stars of the same color. Although there is much scatter in the statistical parallax methods used in their work, brightness variations due to activity warrant consideration because of the photometric parallax relations used to estimate the distances to the stars in our sample. The photometric parallax relation of \cite{bochanski2010} was calibrated with a sample of nearby M dwarfs. Accordingly, the relation is dominated by inactive stars for the early-types and active stars for the late-types. To see what effect this has on our analysis, we made the active stars in our sample 0.5 magnitudes brighter in $M_r$ (or equally fainter in their observed apparent magnitude) when computing their distances. This moved the active stars $\sim25\%$ farther away. In Figure~\ref{fig:actdist} we plot the activity map of dM3 and dM4 stars with the adjusted distances. Compared to Figure~\ref{fig:frac8} and Figure~\ref{fig:frac47}, the adjusted maps are qualitatively very similar, but the streaks of activity extend farther from the mid-plane of the Galaxy.

\begin{deluxetable*}{clccc}  
\tablecolumns{4}
\tablecaption{Effect of Potential Systematics \tablenotemark{i}}
\tablehead{   
  \colhead{Spectral Type} &
  \colhead{Biases/Corrections} &
  \colhead{$a$ (kpc$^{-1}$)} &
  \colhead{$b$ (kpc$^{-1}$)} &
  \colhead{$c$} 
}
\startdata
dM3 & ... & $-0.028\pm_{0.009}^{0.015}$ & $0.035 \pm^{0.017}_{0.020}$&$0.073 \pm0.010$\\[4pt]
 &MisType\tablenotemark{ii}& $-0.029\pm0.007$ & $0.031\pm0.013$ & $0.073 \pm 0.005$ \\[4pt]
  &SpT Shift& $-0.026\pm_{0.009}^{0.012}$ & $0.011\pm0.012$ & $0.058 \pm0.010 $ \\[4pt]
 &$\zeta$& $-0.030\pm_{0.010}^{0.017}$ & $0.026\pm0.019$& $0.075 \pm0.011$\\[4pt]
 &Activity& $0.006\pm_{0.023}^{0.028}$  & $0.055\pm_{0.019}^{0.014}$ & $0.057 \pm0.013$\\[4pt]
   &Activity + SpT Shift& $-0.008\pm_{0.015}^{0.013}$ & $0.018\pm0.015$ & $ 0.054\pm0.010 $ \\[4pt]
  &Activity + $\zeta$& $0.005\pm_{0.020}^{0.026}$ & $0.057 \pm^{0.014}_{0.020}$&$0.057 \pm0.012$\\[4pt]
 &Activity + $\zeta$ + MisType\tablenotemark{ii}& $0.004\pm0.008$ & $0.059 \pm 0.006$&$0.057 \pm 0.004$\\[4pt]
  &Activity + $\zeta$ + SpT Shift& $-0.006\pm_{0.013}^{0.015}$ & $0.022\pm0.016$ & $0.053 \pm 0.010$ \\[4pt]
dM4 &...& $-0.080\pm_{0.012}^{0.014}$& $0.036\pm_{0.038}^{0.022}$&$0.167\pm^{0.017}_{0.015}$\\ [4pt]
&MisType\tablenotemark{ii}& $-0.080\pm 0.017$ & $0.035\pm 0.026   $ &$0.160 \pm0.008 $\\[4pt]\
  &SpT Shift& $-0.075\pm_{0.009}^{0.008}$ & $0.001\pm_{0.009}^{0.006}$ & $ 0.150\pm_{0.014}^{0.013} $ \\[4pt]
 &$\zeta$&  $-0.097\pm^{0.018}_{0.011}$ &$0.011\pm_{0.022}^{0.021}$ &$ 0.175 \pm_{0.016}^{0.016}$\\[4pt]
 &Activity& $ -0.054\pm^{0.037}_{0.028}$&$ 0.071\pm_{0.035}^{0.043}$&$ 0.155 \pm_{0.019}^{0.020}$\\[4pt]
    &Activity + SpT Shift& $-0.061\pm_{0.013}^{0.017}$ & $0.020\pm_{0.018}^{0.023}$ & $0.141 \pm 0.015$ \\[4pt]
  &Activity + $\zeta$& $-0.052\pm_{0.034}^{0.035}$ & $0.068 \pm^{0.044}_{0.041}$&$0.155 \pm^{0.021}_{0.020}$\\[4pt]
 &Activity + $\zeta$ + MisType\tablenotemark{ii}& $-0.054\pm0.033$ & $0.075 \pm 0.023 $&$0.148 \pm0.011$ \\ [4pt]
    &Activity + $\zeta$ + SpT Shift& $-0.068\pm_{0.012}^{0.017}$ & $0.010\pm_{0.014}^{0.024}$ & $0.146 \pm 0.015$ 
\enddata
\label{tab:comb}
\tablenotetext{i}{Parameters from model: $\mathcal{F}  = a |Z| + b(R - R_{\odot}) + c$  ; error bars correspond to 99.73\% confidence interval ($3\sigma$).}
\tablenotetext{ii}{Values correspond to the mean over many realizations of the simulated contamination and the uncertainties correspond to three times the standard deviation of those realizations.}
\end{deluxetable*}

Applying this distance correction to the stars of spectral types dM3 and dM4, we used our quantitative analysis (see Section~\ref{sec:grad}) to determine how the gradients would be affected. For dM3 stars we estimated values of $a = 0.006\pm_{0.023}^{0.028}$ kpc$^{-1}$, $b = 0.055\pm_{0.019}^{0.014}$ kpc$^{-1}$, and $c = 0.057 \pm0.013$.  For dM4 stars we estimated values of $a = -0.054\pm^{0.037}_{0.028}$ kpc$^{-1}$, $b = 0.071\pm_{0.035}^{0.043}$ kpc$^{-1}$, and $c = 0.155 \pm_{0.019}^{0.020}$, where the uncertainties reflect $3\sigma$ confidence intervals. 

Comparing these to the values in Table~\ref{tab:fit}, we see an enhancement of the radial gradient and a weakening of the vertical gradient. As evident in the activity-distance corrected activity maps of Figure~\ref{fig:actdist}, the active stars get moved farther away, which means that instead of having the radial gradient concentrated near the mid-plane it is more prevalent at a range of Galactic heights, making the radial gradient more prominent and weakening the vertical gradient.

\subsection{Combined Effects}\label{sec:comb}

We also combined these potential sources of systematic uncertainties and applied our methods. In Table~\ref{tab:comb} we list the combinations and the corresponding determinations for the activity distribution gradients. In Figure~\ref{fig:actzetadist} we show the activity maps for dM3 and dM4 stars with corrected distances according to activity and metallicity. When combining the corrections, the enhancement of the radial gradient due to the activity distance correction dominates over the other biases. The weakening of the radial gradient due to the spectral type shifting is stronger than the similar effect from correcting the potential metallicity bias as can be seen by comparing the combinations of these with the activity correction. Regardless, for reasons stated above, the spectral types biases are unlikely to be prominent in the data and the most relevant corrections would be the activity and metallicity analyses. We conclude then that for both dM3 and dM4 stars the radial gradient is significant beyond the $3\sigma$ confidence level.

\begin{figure}[htbp]
\centering
\includegraphics[width=0.46\textwidth,clip=true,angle=0]{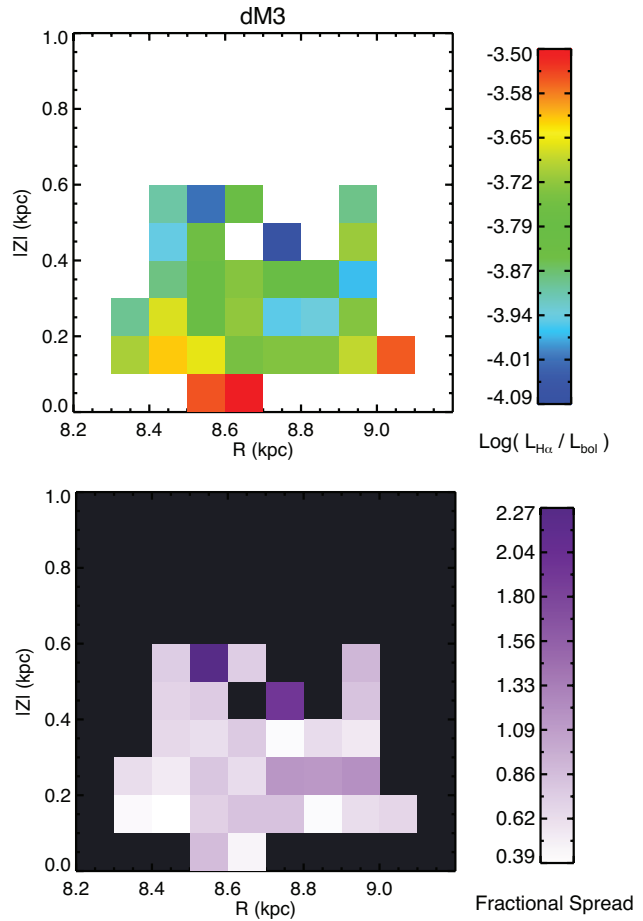}
\caption[]{ Top - dM3 activity level as a function of Galactocentric radius and distance from the Galactic plane. The color corresponds to the median level of activity, $L_{\rm{H}\alpha} / L_{\mathrm{bol}}$ in each bin. Redder colors correspond to bins that are more active. Bottom - Map of the spread of data, width of middle 50\%, divided by the median activity level for each bin corresponding to the activity level map of the top panel. Sometimes the width of the spread is greater than the median value of the activity in the spatial bin. Lighter shades correspond to narrower distribution. Bins are 100 pc by 100 pc.}
\label{fig:m3_Ha}
\end{figure}

\section{Degree of Activity}\label{sec:deg}

In the maps of the previous section, we have shown how the incidence of activity varies spatially and by spectral type. Additionally, we also examined how the level of activity, $L_{\mathrm{H}\alpha}/ L_{\mathrm{bol}}$, varies with Galactic position. We estimated $L_{\mathrm{H}\alpha}/ L_{\mathrm{bol}}$ by using the $\chi$ factor of \cite{walkowicz2004} to convert our measured equivalent widths  in H$\alpha$ to $L_{\mathrm{H}\alpha}/ L_{\mathrm{bol}}$. We included only those stars that were deemed `active' following the criterion of Section~\ref{sec:data} and used the same binning analysis as Section~\ref{sec:map} except that we set the minimum number of active stars to five in each bin to be included in the map. The number of stars in each spectral type used for this part of the analysis is shown in Table~\ref{tab_act} under the `active' column. In Figure~\ref{fig:m3_Ha}, we show an example of this analysis for dM3 dwarfs. The top panel shows the median of the level of activity, $L_{\mathrm{H}\alpha}/ L_{\mathrm{bol}}$, for the stars in each bin with redder colors being more active. The bottom panel shows the total range spanned from the 25th quartile to the 75th quartile divided by the median of the activity level. There appears to be a decrease in median activity level away from the Galactic plane, as was first shown by \cite{west2008}. There were no hints of radial trends in our maps. However, the two dimensional plots are inconclusive with high uncertainties and likely require a larger sample size to make definitive arguments. Consequently, we did not apply our quantitative two dimensional analysis (see Section~\ref{sec:grad}) to the activity level maps. Although, there did not appear to be a radial gradient in the level of activity, among only the active stars, it is clear from Figure~\ref{fig:m3_Ha} that at larger Galactocentric radii ($R > 8.6$ kpc) there are many more bins with active stars than there are at smaller Galactocentric radii ($R < 8.6$ kpc), leading to the significant radial gradient in the activity distribution (see Table~\ref{tab:comb}).

\section{Discussion and Summary}\label{sec:discuss}

The large sample of M dwarfs from the SDSS has allowed us to investigate the spatial distribution of M dwarf properties in the Galactic radial domain, $R$, in addition to the Galactic vertical domain, $Z$, above and below the plane. We reproduced and refined previous results concerning the decline in magnetic activity (as traced by H$\alpha$ emission) with increasing absolute distance from the Galactic plane. We examined this more closely by considering each of the spectral types individually. For the early types (dM0--dM2) there was little evidence of any spatial correlations with magnetic activity fraction; almost all of these stars were observed to be inactive. By contrast, more than 50\% of dM7 stars were seen to be active. This is likely a reflection of the mean age of the stellar disk being much greater than the active lifetime of the early-type stars ($\lesssim$1.5 Gyr ; \citealt{west2008}) and the active lifetime for dM7 stars ($\sim$8 Gyr ; \citealt{west2008}) being much longer than the average age of the local disk.
 
 In Figures \ref{fig:frac8} and \ref{fig:frac47}, within the maps for types dM3--dM7, we see many active stars and a strong correlation between activity and absolute vertical distance from the Galactic mid-plane with vertical gradients inconsistent with zero. The maps for dM6 and dM7 spectral types are particularly striking because of the large drop off in activity over a short distance. The vertical gradients in the activity distribution trace the drop in mean stellar age with absolute vertical distance from the mid-plane. The large gradients we have measured however, may be systematically high (see Section~\ref{sec:dist}).

We also showed evidence for a potential radial trend in activity, strongest in the analysis of stars of spectral type dM3 (inconsistent with zero at $3\sigma$ confidence level) but also apparent in the analysis of stars of spectral type dM4. We tested the robustness of this result against potential systematics from spectral misclassification and/or luminosity effects related to the activity/metallicity of the stars. Our test of spectral misclassification showed that although statistical mis-typing could change the exact measure of the gradients, it did not erase the radial trend in the data. We also considered a systematic misclassification as a function of the distance to a given star. Correcting for this kind of bias significantly reduced the radial gradients, especially for the dM4 stars. However, because of the careful visual spectral typing procedure of Paper I, the types should be accurate to 0.5 subtypes and the misclassification should be minimal and resistant to the systematic effects that could lead to the distance related bias.

In testing for effects related to metal content, we found that the underlying metallicity distribution of the sample did systematically boost the apparent radial gradient in the activity distribution. In applying a correction, the trend disappeared for dM4 stars but persisted for dM3 stars. We then tested an additional luminosity correction for `active' vs `inactive' stars and showed that it greatly enhanced the significance of the radial trends for both dM3 and dM4 stars and overcame the effects of the metallicity correction, even for dM4 stars (see Table~\ref{tab:comb}).

One possible explanation for the radial trend is that it is a result of the past star formation history in the Galactic disk. The denser regions near the Galactic center produced more stars early on compared to the outer parts of the disk, resulting in a negative radial stellar age gradient. The observed trend in activity could be indicative of this gradient in average age for the dM3 dwarfs. 

Other effects may also play an important role. The spiral galaxy N-body simulations by \cite{loebman2011}, which included radial migration in their attempt to reproduce the Milky Way showed a negative radial age gradient in the disk of $\sim$1 Gyr kpc$^{-1}$, similar to the observations of GCS \citep{nordstrom2004, casagrande2011}. The observed trend is likely a combination of these mechanisms operating over the course of the evolution of the Milky Way. 

Age differences of 1 Gyr at different Galactocentric radii are unlikely to affect the activity distribution of late-type stars (dM5--dM7) because their lifetimes are so long compared to the local age of the disk. Additionally, the earliest type (dM0--dM2) stars are mostly already past their active lifetimes so their distribution is also flat. However, for the mid-type M dwarfs (dM3--dM4), a difference of 1 Gyr is a large fraction of their active lifetimes, allowing for the observed distribution of activity fractions we have found here in the distributions of SDSS M dwarfs.

Another way to interpret the maps of Figure~\ref{fig:frac8} and Figure~\ref{fig:frac47} is as a time series for a single star type going from early times to late times when going from the dM7 map to the dM0 map. Starting with the the map for dM6/dM7 stars, the stars are mostly active with some older stars farther from the mid-plane starting to turn off their activity as a consequence of a vertical age gradient. With an underlying radial age gradient, the oldest stars closer to the Galactic center also turn off first as time passes, decreasing the median activity fraction as we saw in the maps for dM4 stars. As in that map, the transition region of zero activity to moderate activity develops the sloped contour from the confluence of the vertical and radial age gradients. Over time that region shifts away from the Galactic center as the stars farther out pass their activity lifetimes as seen in the map for dM3 stars. As the population ages the activity fraction bottoms out close to 0\% with most stars being inactive as in the maps for dM0/dM1 stars.

Our measurement of the activity distribution catches the local population of M dwarfs at a time when the ages of the local populations of intermediate M dwarfs, dM3 and dM4 spectral types, bridges the transition between `active' and `inactive' stars. In a couple Gyrs this will apply to the later stars of spectral types dM5, dM6 and dM7.

In order to get a better sense of potential age trends using activity and distinguish between different evolutionary models of the MIlky Way, it is imperative that we concretely establish the functional relationship between age and magnetic activity in M dwarfs; such studies in addition to larger and more distant samples will allow for a better examination of the structure and evolution of the Galaxy.

\section*{Acknowledgments}

J.S.P would like to thank the referee for strengthening the results of this work as well as Elisabeth R. Newton and Saurav Dhital for useful comments in the preparation of this work. J.S.P. acknowledges the Paul E. Gray fund in providing monetary support for the Undergraduate Research Opportunities Program at MIT. J.S.P acknowledges support from the National Science Foundation Graduate Research Fellowship under Grant No. (DGE-1144469). A.A.W acknowledges funding from NSF grant AST-1109273. AAW also acknowledges the support of the Research Corporation for Science Advancement's Cottrell Scholarship

Funding for the Sloan Digital Sky Survey (SDSS) and SDSS-II has been
provided by the Alfred P. Sloan Foundation, the Participating
Institutions, the National Science Foundation, the U.S. Department of
Energy, the National Aeronautics and Space Administration, the
Japanese Monbukagakusho, and the Max Planck Society, and the Higher
Education Funding Council for England. The SDSS Web site is
http://www.sdss.org/.

The SDSS is managed by the Astrophysical Research Consortium (ARC) for
the Participating Institutions. The Participating Institutions are the
American Museum of Natural History, Astrophysical Institute Potsdam,
University of Basel, University of Cambridge, Case Western Reserve
University, The University of Chicago, Drexel University, Fermilab,
the Institute for Advanced Study, the Japan Participation Group, The
Johns Hopkins University, the Joint Institute for Nuclear
Astrophysics, the Kavli Institute for Particle Astrophysics and
Cosmology, the Korean Scientist Group, the Chinese Academy of Sciences
(LAMOST), Los Alamos National Laboratory, the Max-Planck-Institute for
Astronomy (MPIA), the Max-Planck-Institute for Astrophysics (MPA), New
Mexico State University, Ohio State University, University of
Pittsburgh, University of Portsmouth, Princeton University, the United
States Naval Observatory, and the University of Washington.

\bibliographystyle{apj}
\bibliography{ms}

\end{document}